\documentclass[prd,aps,preprintnumbers,superscriptaddress,onecolumn]{revtex4}

\usepackage{graphicx} 

\begin{document}

\title{Search for Sterile Neutrinos with a Radioactive Source at Daya Bay}

\author{D.A.~Dwyer}\affiliation
        {\it Kellogg Radiation Laboratory and Physics Department,
Caltech, Pasadena, CA 91125, USA}
\author{K.M.~Heeger}\affiliation
        {\it University of Wisconsin, Department of Physics, Madison, WI 53706, USA}
\author{B.R.~Littlejohn}\affiliation
        {\it University of Wisconsin, Department of Physics, Madison, WI 53706, USA}
\author{P.~Vogel}\affiliation
        {\it Kellogg Radiation Laboratory and Physics Department,
Caltech, Pasadena, CA 91125, USA}

\date{\today}

\begin{abstract}
The far site detector complex of the Daya Bay reactor experiment is proposed as a location to search for sterile neutrinos with $\geq$~eV
mass.  Antineutrinos from a 500~kCi $^{144}$Ce-$^{144}$Pr $\beta$-decay source (${\Delta}Q=$ 2.996~MeV) would be detected by
four identical 20-ton antineutrino targets.  The site layout allows flexible source placement; several specific source locations are
discussed.  In one year, the 3+1 sterile neutrino hypothesis can be tested at essentially the full suggested range of the parameters
$\Delta m^2_{\rm new}$ and $\sin^2 2\theta_{\rm new}$ (90\% C.L.).  The backgrounds from six nuclear reactors at $\geq$1.6~km distance are
shown to be manageable.  Advantages of performing the experiment at the Daya Bay far site are described.
\end{abstract}

\maketitle

\section{Introduction}

Neutrino mass and mixing are usually incorporated in the generalization of the Standard Model by assuming that the three neutrino states of a
given flavor, $\nu_e$, $\nu_{\mu}$ and $\nu_{\tau}$, are superpositions of the three mass eigenstates $\nu_i$,
\begin{equation} \nu_{\ell} = \Sigma_{i=1}^3 U_{\ell, i} \nu_i ~.
\end{equation} 
Here $U_{\ell, i}$ is the unitary 3$\times$3 neutrino mixing matrix. This assumption makes it possible to consistently describe most solar, atmospheric, reactor and accelerator neutrino experiments. The values of the mixing angles in $U_{\ell, i}$ as well as of the mass square differences $\Delta m_{21}^2 \equiv \Delta m_{sol}^2$ and $|\Delta m_{31}^2| \equiv \Delta m_{atm}$ can be deduced from analysis of these data~\cite{PDG}.

However, several recent experiments indicate that this picture might be
incomplete, although the statistical significance is limited. See~\cite{Giunti2011} and references therein.  They
suggest that one or more sterile neutrinos, which weakly couple to the
active neutrinos, might exist. In particular, the ``reactor anomaly''~\cite{Mention}, based on the re-evaluation of the nuclear reactor
$\overline{\nu}_e$ flux~\cite{Mueller}, leads to the conclusion that the
$\overline{\nu}_e$ produced in the reactor core oscillate into some sterile
neutrino species at distances of less than $\sim$10 m  from the
reactor core. This would reduce the active $\overline{\nu}_e$ flux observed
by experiments at distances greater than 10 m from the reactor.

With the modified reactor flux model of Ref.~\cite{Mueller}, the
simplest 3+1 sterile neutrino model was used to analyze the existing
results from reactor experiments, GALLEX and SAGE calibration sources,
and MiniBooNE~\cite{Mention}.  That work included the reanalysis of
the MiniBooNE experiment~\cite{Giunti2010} and the results of the ILL
reactor experiment~\cite{Kwon}.  The resulting best fit sterile
neutrino oscillation parameters are $|\Delta m^2_{\rm new}|= 2.35\pm0.1$
eV$^2$ (68\%~C.L.) and $\sin^2(2\theta_{\rm new})= 0.165\pm0.04$ (68\%~C.L.). The region of these parameters compatible at 95\% C.L. with all the
analyzed experiments is constrained by $|\Delta m^2_{\rm new}|>$ 1.5
eV$^2$ and $\sin^2(2\theta_{\rm new})=0.17\pm0.09$. A recent analysis of short-baseline neutrino oscillation data in the framework
of 3+1 neutrino mixing including the update of MiniBooNE antineutrino data and
the MINOS results~\cite{Giunti2011} yields best-fit values from around $\Delta m^2_{41} \approx$ 1~eV$^2$ up to $\approx 5.6$~eV$^2$, where $\Delta m^2_{41}$ corresponds to the new mass splitting $\Delta m^2_{\rm new}$ in the specific 3+1 model. 

To test the hypothesis of a short distance oscillation into
a sterile neutrino state ideally one would like to place a
$\overline{\nu}_e$ detector near the distance
\begin{equation} \label{eq:oscLength}
L_{optim} = \frac{L_{osc}[{\rm m}]}{2} = 
1.24 \frac{E_{\overline{\nu}_e}[{\rm MeV}]}{\Delta m^2_{\rm new} [{\rm eV}^2]} 
\end{equation}
and observe the $L/E_{\overline{\nu}_e}$ variation of the observed signal.  

Electron antineutrinos are produced in radioactive decays. Beta decays in the fission fragments of nuclear fuel make nuclear reactors a convenient source of $\overline{\nu}_e$. From the the discovery of the $\overline{\nu}_e$ by Reines and Cowan in 1960 to the recent observation of $\overline{\nu}_e$ disappearance at KamLAND~\cite{KamLAND2003}, reactor antineutrinos have played an important role throughout the history of neutrino physics. For reactor  ${\overline{\nu}_e}$ the maximum signal is near
$E_{\overline{\nu}_e}\!\simeq$4~MeV\@.  Using the most probable $\Delta
m^2_{\rm new}$ given above, Eq.~\ref{eq:oscLength} gives the optimum
distance of $\sim$2~m.  This length scale makes commercial power reactors with typical reactor core dimensions of 3-5~m not suitable for the search for sterile neutrinos. The size of their cores would smear the oscillation pattern. See also Ref.~\cite{Yasuda}. Smaller, more compact research reactors partially avoid this problem but placing detectors at distances of 1-5~m from a reactor core is generally a challenge. A number of efforts are currently underway to investigate the feasibility of sterile neutrino searches at research reactors~\cite{DANSS, NUCIFER,Bowden}.

Avoiding the issue of smearing and extended reactor cores, another approach is to use a ``point-like'' radioactive antineutrino source. It was recently proposed by Ref.~\cite{Cribier} to place
a strong radioactive source of antineutrinos  in the center of one of the large liquid scintillation detectors such as KamLAND,
Borexino, or SNO+.  To detect the $\overline{\nu}_{e}$ from the source, the Q-value must be above the
detection threshold of inverse neutron beta decay ($>$1.805~MeV)\@.  A
variation of the interaction rate versus the distance from the
antineutrino source, and with the corresponding energy, would be
evidence of oscillation into a sterile neutrino state.  The proposed
source is $^{144}$Ce, with an intensity of 50~kCi
(1.85$\times$10$^{15}$~Bq).  This decays into the unstable daughter
$^{144}$Pr which, in turn, decays into the stable $^{144}$Nd with the
Q-value of 2.996~MeV\@.  The $^{144}$Pr decay produces antineutrinos
above the 1.8~MeV threshold for inverse neutron beta decay.  The
half-life of $^{144}$Ce is 285 days and of its daughter $^{144}$Pr only
17.3 minutes, so that the latter decay remains in equilibrium at all
times.  Since $A=$ 144 is near the top of the fission yield curve, the isotope $^{144}$Ce is contained in considerable quantities of several percent in the fission  fragments of spent nuclear fuel. We will not discuss here the
challenges of extracting the radioactive $^{144}$Ce from the used
reactor fuel rods and preparing the suitable source. Many PBq of Ce are typically contained in one fuel rod at full burn-up. This makes it feasible to obtain sufficient $^{144}$Ce during reprocessing of spent nuclear fuel to prepare such a source. 

While the test proposed in Ref.~\cite{Cribier} is possible, placing a strong and heavily-shielded radioactive source in the middle of very clean
detectors such as KamLAND, SNO+, or Borexino is a challenging technical problem. Spatial constraints in the access ports to the inner regions of these detectors may also limit the amount of gamma or neutron shielding that can be practically used around such a source.  It has been realized~\cite{ReactorFNAL} that the unique geometry of the Far Hall of the Daya Bay reactor $\theta_{13}$ experiment~\cite{DayaBayexperiment,DayaBay} provides an opportunity for an oscillation measurement with multiple detectors over baselines of $\sim$1-10~m using a radioactive $\overline{\nu}_{e}$ source or spent nuclear fuel outside the antineutrino detectors. In this paper we propose an alternative source experiment by placing a $\overline{\nu}_{e}$ source such as $^{144}$Ce source in the space between the four antineutrino detectors in the Far Hall of the Daya Bay experiment~\cite{DayaBayexperiment,DayaBay}.  The unique geometry of the detector arrangement in the Daya Bay Far Hall allows one to place an antineutrino source with $\ge$35~cm of shielding as close as $\sim$1.3~m from the active detector region while staying outside the antineutrino detectors in a water pool that provides convenient shielding and source cooling.

\section{A Source Experiment at Daya Bay}

\subsection{Antineutrino Detectors at Daya Bay}
The Daya Bay reactor experiment is a next-generation reactor experiment designed to make a high-precision measurement of the neutrino mixing angle $\theta_{13}$ using antineutrinos from the Day Bay reactor complex near Hong Kong, China~\cite{DayaBayexperiment}. The Daya Bay experiment uses three underground experimental halls at distances ranging between 350-2000~m from the nuclear power plant to measure the disappearance of $\overline{\nu}_e$ as a function of distance from the reactor source.
Eight antineutrino detectors are placed at
three underground locations near the Daya Bay reactor facility.  Two
detectors are deployed side-by-side $\sim$360~m from the original Daya Bay reactor
cores in the Daya Bay Near Hall while two other detectors are placed $\sim$500 m from the four newer Ling
Ao reactor cores in the Ling Ao Near Hall.  In addition, four detectors are placed at a distance of $\sim$1600-2000~m
 from all 6 reactor cores in the Daya Bay Far Hall.  In the absence of sterile neutrino oscillations the near detectors provide a precise
measurement of the unoscillated antineutrino flux, while the far
detectors will measure the oscillation parameter and neutrino mixing angle
$\theta_{13}$ from the suppression of the $\overline{\nu}_e$ and the distortion of its energy spectrum.  Any oscillation into sterile species largely cancels in the measured ratio of event rates and spectra between the near and far detectors and hence does not enter the $\theta_{13}$ measurement. 

\begin{figure}[htb]
\includegraphics[width=0.8\textwidth,angle=0]{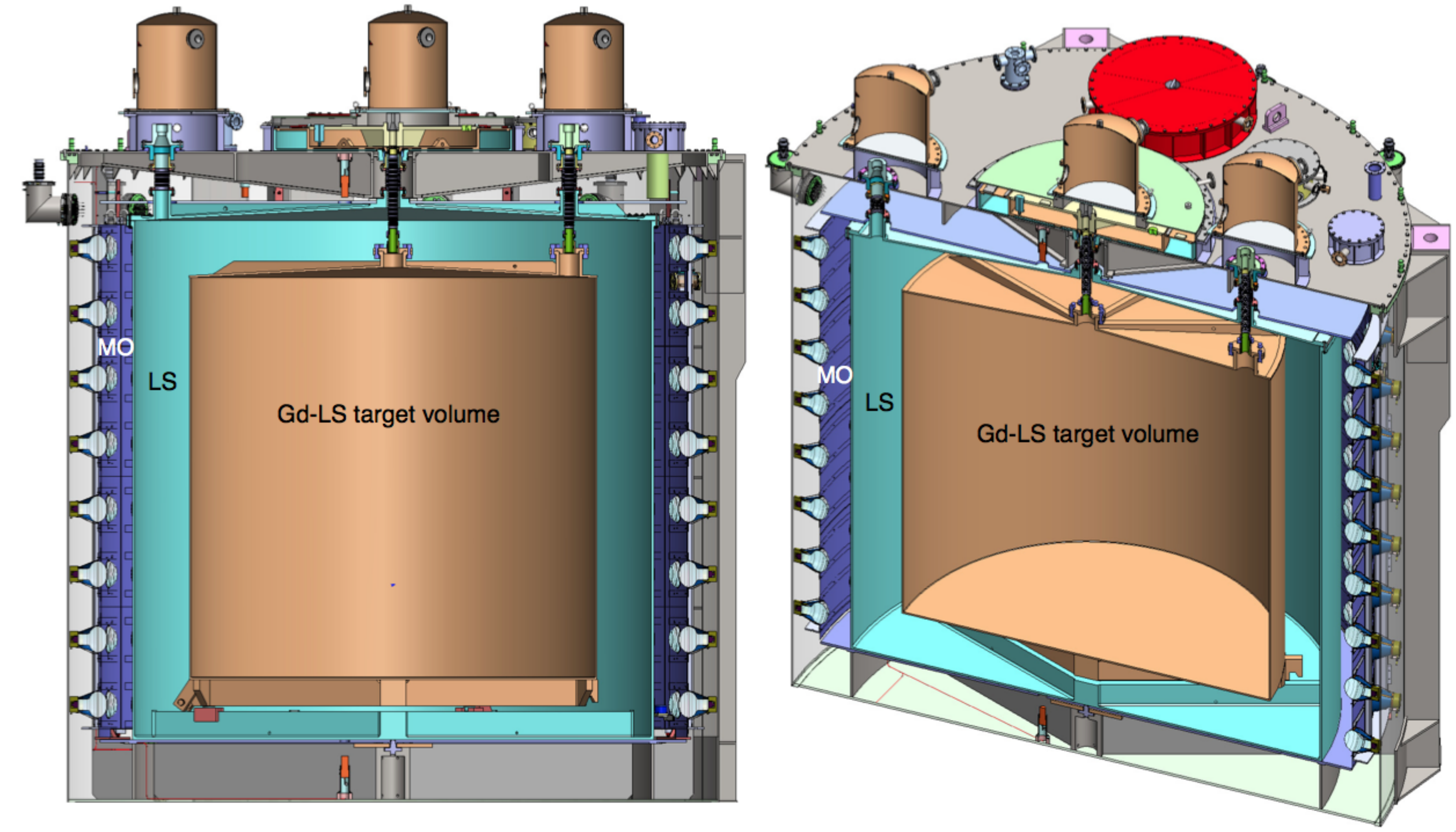}
\caption{Schematic cross-section model of a Daya Bay antineutrino detector showing the three nested, concentric volumes of Gd-loaded liquid scintillator, pure liquid scintillator (LS), and mineral oil (MO). Calibration and instrumentation systems are shown on top of the detectors.  Figures adapted
  from Ref.~\cite{DayaBay, DayaBayTAUP2011}.}
\label{fig:ad}
\end{figure}

Each Daya Bay detector is composed of three nested
volumes, separated by transparent acrylic cylinders as shown in
Fig.~\ref{fig:ad}.  The inner most volume is a 20-ton Gd-loaded
liquid scintillator antineutrino target cylinder with a diameter and height of $\sim$3~m.  The
next volume is a 20-ton cylindrical region of pure liquid scintillator (LS) with an outer diameter of $\sim$4~m, a thickness of $\sim$50~cm and a height of 4~m. It is designed to improve the efficiency for capture of
$\gamma$-rays produced by antineutrino interactions in the target
region.  Outermost with a height and diameter of about 5~m is a 40-ton buffer shell of inactive mineral oil (MO) which attenuates external background radiation.  Some 192
8''-photomultipliers are mounted  in this region along the vertical detector walls to collect the
scintillation light produced by antineutrino interactions in the
target.  A cylindrical stainless steel tank provides containment and
support.

The underground hall at the Daya Bay far site houses four antineutrino detectors. The floor plan and layout of this experimental hall
are shown in Fig.~\ref{fig:FarHallfig2}.  Four antineutrino detectors are arranged in a square
within an octagonal water pool. The centers of the antineutrino detectors are separated by about 6~m. The pool is instrumented with photomultipliers for use as a cosmic muon veto.  The water pool is $\sim$10~m deep, providing an additional 2.5~m of water veto above and below
each antineutrino detector.  The pool structure and concrete floor are designed to carry the load of the antineutrino detectors. Each detector weighs about 110~t when filled with liquids. The experimental hall has an overhead crane capable of
lifting 125~ton loads. Access to the underground experimental halls is through a tunnel large enough to allow the transport of the 5~m high and 5~m wide detectors. For a description of the Daya Bay experiment see
Ref.~\cite{DayaBay, DayaBayTAUP2011}.

\subsection{Source Locations}

For the source experiment proposed here we assume a 500~kCi $^{144}$Ce source, similar in design to the one described in Ref.~\cite{Cribier}.  For a spherical source of that strength the radius will be about 8~cm.  The spatial extent of the source material will be small compared to the neutrino oscillation lengths considered here. In order to shield the $\gamma$ radiation from the 1\% branch to the $1^-$ excited state at 2.185~MeV in $^{144}$Nd, as well as the Bremsstrahlung accompanying the $\beta$-decay, the compact source will be surrounded by some 35~cm of shielding; 33~cm of W and an additional 2~cm of Cu. For a 50~kCi source that will reduce the 2.185~MeV $\gamma$-rays by a factor of 2$\times$10$^{-10}$ to $\sim$4~kBq. More shielding can be used for a higher-activity source, and for a ten times stronger source of 500~kCi one would need to add about 3~cm of W to achieve another order of magnitude in background reduction. The source locations proposed here allow the use of additional shielding if necessary. The physical outer dimensions of the antineutrino detectors and the source geometry define the minimum distance between the source material and the active antineutrino detector region. When the source including its shielding is placed directly in contact with the outer stainless vessel of the antineutrino detectors the minimum distance between the active source material and the active detector region is about 
$\sim$1.3~m. Since the source dimensions have no stringent limitations in the water pool, additional shielding material to mitigate gammas or neutrons from the source material can be added. Increasing the source shielding and its distance from the active detector region will also decrease the solid angle seen by the detector.  At this specified source activity thermal heat is an issue. When the source is outside the pool the source must be actively cooled, since it produces 7.5~W/kCi.

\begin{figure}[htb]
\includegraphics[width=0.8\textwidth,angle=0]{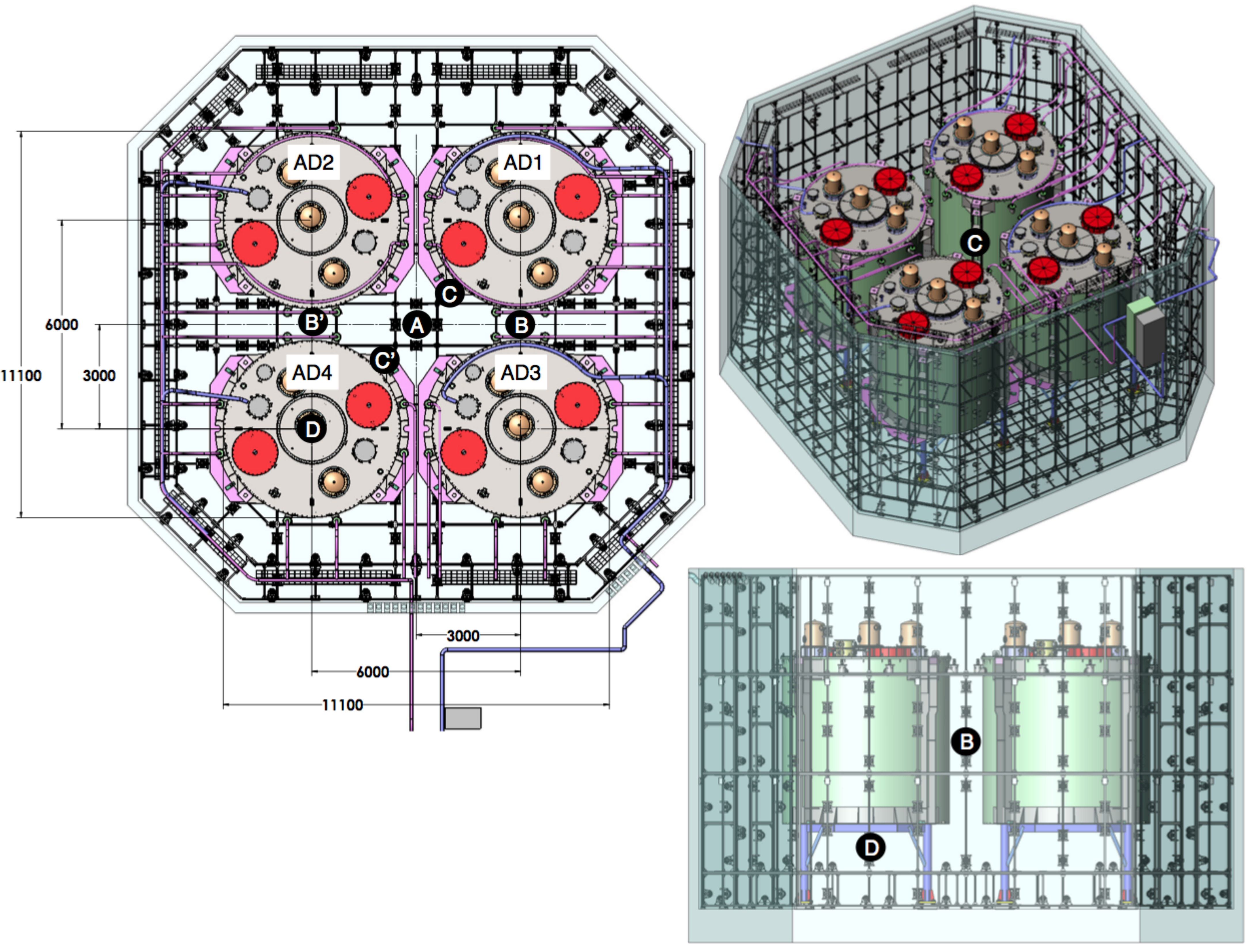}
\caption{Model of the four antineutrino detectors (AD1-4) in the Daya Bay Far Hall. Left: Top view of the
 Far Hall with water pool (octagonal frame), four antineutrino detectors
  (grey cylinders) on their support stands (pink), and water cosmic ray veto photomultipliers and support structure (small
  black features).  {\em A}, {\em B}, {\em C}, and {\em D} mark
  possible antineutrino source locations. Positions {\em B'} and {\em C'} are symmetric to  {\em B} and {\em C} and can be used as cross-checks and for systematic studies. Detector dimensions are given in mm. Right: ISO and side view of the Far Hall.  {\em A}, {\em B}, {\em C} are all at half-height of the antineutrino detectors, {\em D} is directly below it. Figures adapted from Ref.~\cite{DayaBay, DayaBayTAUP2011}.}
\label{fig:FarHallfig2}
\end{figure}

We have explored several possibilities of placing the intense
$^{144}$Ce radioactive source in the water between the four
detectors in the Daya Bay Far Hall. In particular, we have made a detailed study of three
positions: In the center, i.e. equally distant from all four
detectors (position $A$), on one of the sides equally spaced between the centers of
two of the detectors (position $B$), and at the closest point to one of the antineutrino detectors (position $C$). Given the symmetry of the detector arrangements, positions $B'$ and $C'$ have the same physics capabilities as positions B and C\@. With a sufficiently strong source one can consider swapping the source between positions $B$ and $B'$ or $C$ and $C'$ respectively. This can provide a cross-check and help understand detector systematics as well as backgrounds.  Swapping allows us to compare the response of different antineutrino detectors to the same source over the same baseline.  In all cases, the source is placed in the plane defined by the vertical center of the antineutrino target volumes that is at half-height of the antineutrino detectors. Fig.~\ref{fig:FarHallfig2} illustrates the locations of these source positions.

Another possibility is to place the source directly below one of the antineutrino detectors as shown in position $D$ in Figure~\ref{fig:FarHallfig2}. This position would allow one to make a measurement of the $\overline{\nu}_{e}$ oscillation probability along the vertical axis of one cylindrical neutrino detector and take advantage of the symmetry of the detector. In this case spherical contour lines of equal distance around the source more naturally divide the inner target volume of the detector immediately above the source as the radius vector of these contours aligns with the vertical symmetry axis of the detector. A source position directly below the antineutrino detector is preferable over a source position above the detector as calibration systems and instrumentation block access to the detector top and increase the minimum distance between the source and the detector volume. A qualitative comparison of the source positions is given in Table~\ref{tab:sourcepositions}.

\begin{table}[htbp]
   \centering
   \begin{tabular}{c | l} 
   \hline
Source Position & Characteristics (range of baselines, event rate)\\
\hline
\hline
A & Same baselines to all four detectors. \\
 & Smallest range of baselines within active detector region. \\
 & Lower statistics. \\
 \hline
B & Samples 2 principal baselines between source and active detector region. \\
 & Highest statistics, two detectors per baseline. \\ 
 & Uses two-fold symmetry of detector configuration. \\
 \hline
C & Samples widest range of baselines with 3 principal distances.\\
& Lower statistics. \\
 \hline
D  & Lower statistics, mostly collect statistics from one detector.\\
 & Does not use symmetry of Far Hall.\\
 & Utilizes symmetry of detector. \\
 \hline
   \end{tabular}
   \caption{Characteristics of source positions and qualitative comparison. Due to the high statistics, position $B$ is one of the favorite positions for the proposed source measurement. }
   \label{tab:sourcepositions}
\end{table}

\begin{figure}[htb]
\includegraphics[width=0.9\textwidth,angle=0]{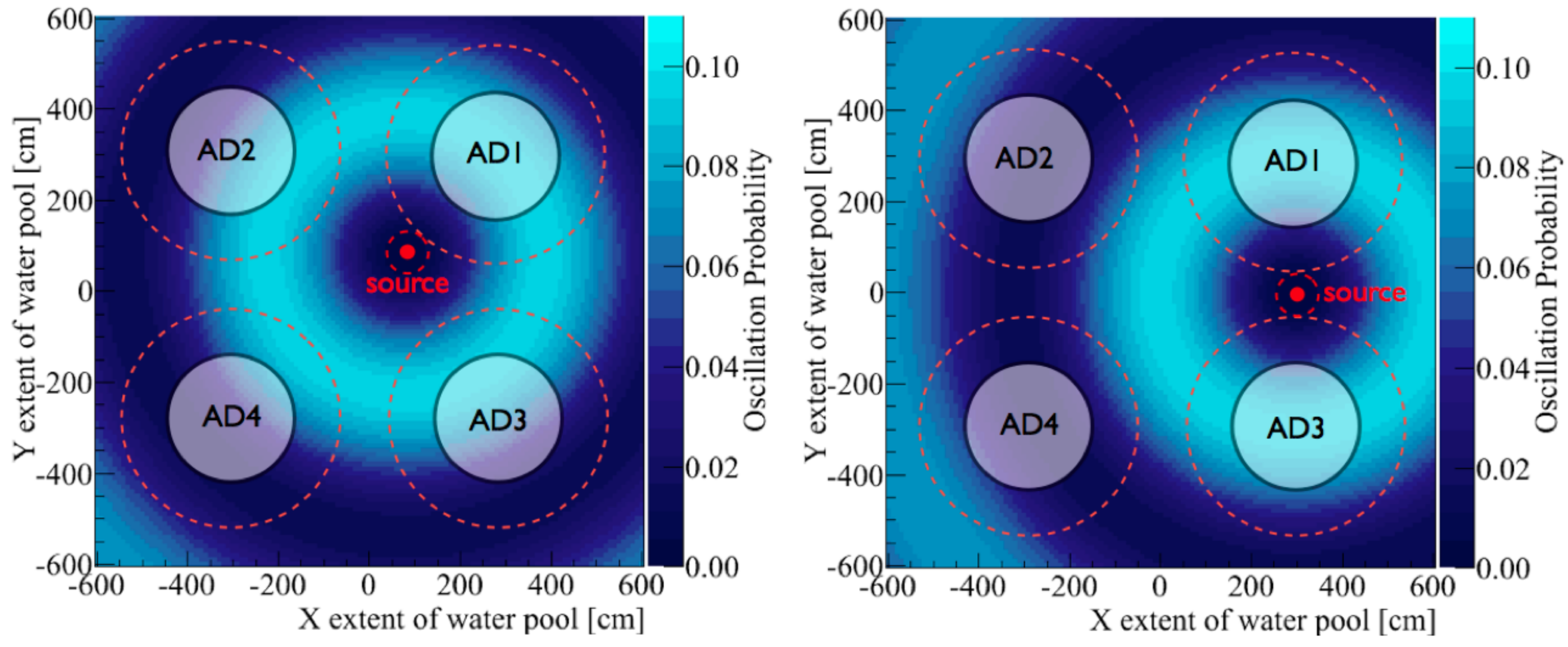}
\caption{Illustration of sterile neutrino oscillation in the Daya Bay Far Hall and top view of the geometric arrangement of the antineutrino detectors and the $\overline{\nu}_{e}$ source. Left: Source at position {\em C}. Right: Source at position {\em B}. Figures show an overlay of the positions of the active regions of the antineutrino detectors with the disappearance probability for $\overline{\nu}_{e} \rightarrow \nu_{s}$ oscillation with $\Delta m^2_{41}$=1~eV$^2$ and sin$^22\theta_{14}$=0.1 into sterile species.  Active regions of the source and detectors are shown in solid red and grey color respectively, while the physical outer dimensions of the source and detectors are indicated by the dashed lines.  }
\label{fig:OscProbfig3}
\end{figure}

Figure~\ref{fig:OscProbfig3}  illustrates the oscillation pattern in the Daya Bay Far Hall for $\overline{\nu}_{e} \rightarrow \nu_{s}$ oscillation into sterile species for source positions $C$ (left) and $B$ (right). For the purpose of this illustration we assume a sterile oscillation with $\Delta m^2_{14}$=1~eV$^2$ and sin$^22\theta_{14}$=0.1\@. The figure shows the color-coded disappearance probability and the positions of the active regions of the antineutrino detectors. The active regions of the source and detectors are shown in solid red and grey color respectively, while the physical outer dimensions of the source and detectors are indicated by the dashed lines. 

The detector configuration in the Daya Bay Far Hall provides a unique geometry and environment for an experiment with a strong antineutrino source. Multiple source positions outside the containment vessels of the antineutrino detectors are possible. The symmetric arrangement of the detectors allows for cross-checks, systematic comparisons, and swapping. The symmetry of four identical detectors allows additional tests of systematic errors. The existing water pool provides a thermal bath for a hot source, free shielding, and few spatial constraints on the size and location of the source.  Infrastructure exists for the transport and lifting of heavy objects up to 120 tons. The placement of the antineutrino source on the outside of the detectors avoids complicated issues of cleanliness and all source locations described above allow the addition of extra shielding to further suppress $\gamma$ or neutron backgrounds. For position $C$ and $C'$ respectively this increases the minimum distance between the active source material and the active detector region by the amount of the shielding. For source position $B$ and $B'$ we are simply limited by the available space between the detectors.

\subsection{Signature of $\overline{\nu}_{e}$ Oscillation into Sterile Neutrinos}

\begin{figure}[htb]
\includegraphics[width=.6\textwidth,angle=0]{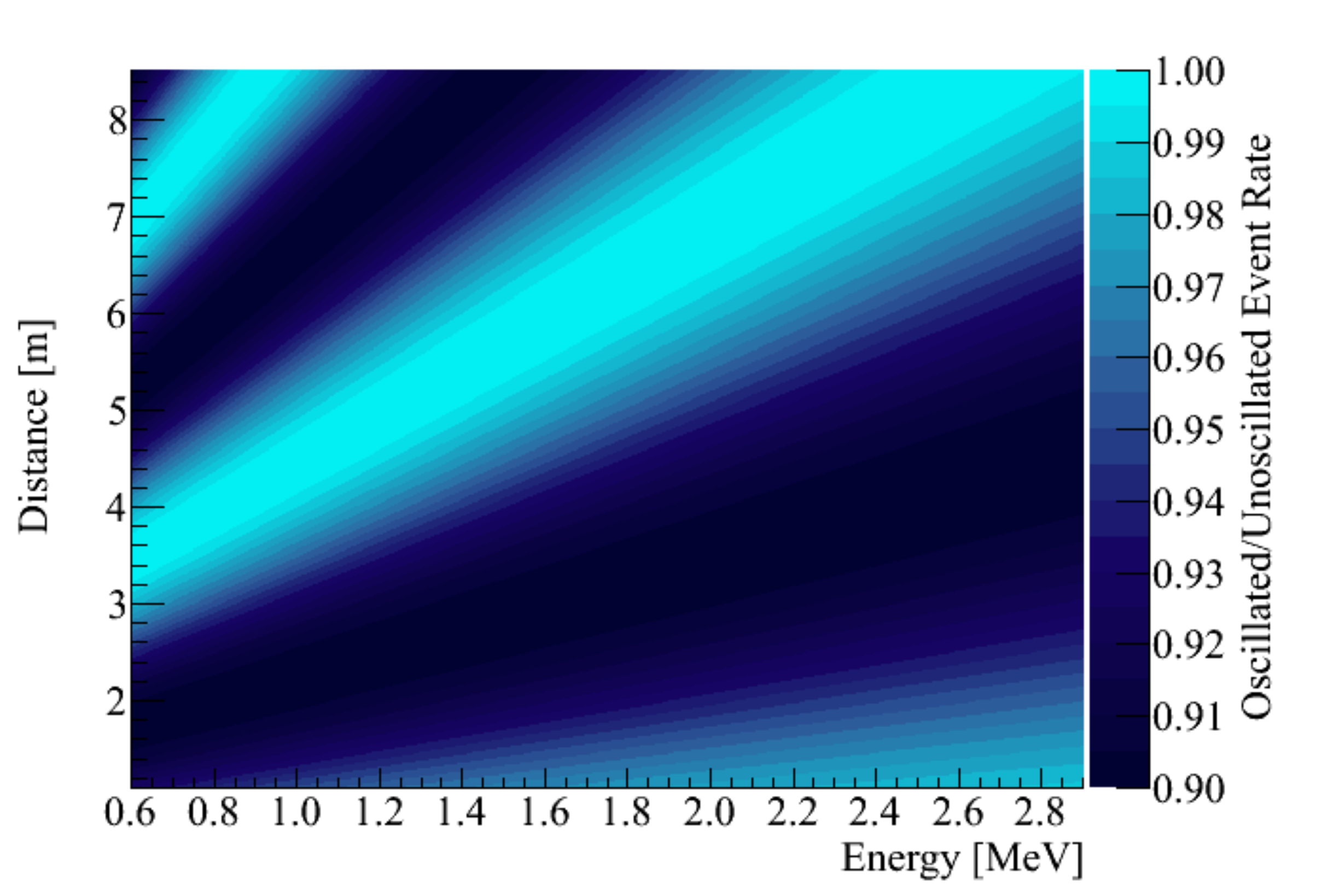}
\caption{Example oscillation effect as a function of energy and distance from the $\overline{\nu}_{e}$ source for $\Delta m_{41}=1$~eV$^2$ and sin$^2\theta_{14}$=0.1\@.  The figure shows the ratio of the oscillated over the unoscillated event rate. }
\label{fig:RateOscfig5}
\end{figure}

In our proposed source experiment a spectrum of $\overline{\nu}_{e}$ is emitted from an almost point-like source. The antineutrinos travel a range of distances determined by the layout of the detector arrangement and the cylindrical symmetry of the detectors before they interact in the Daya Bay antineutrino detectors. The typical distance traveled from the source to the detector ranges from $\sim$1.5-8~m. The energy spectrum is determined by the $^{144}$Ce source. The energy and distance traveled determine the oscillation probability of the $\overline{\nu}_{e}$ flux. Figure~\ref{fig:RateOscfig5} shows the effect of $\overline{\nu}_{e} \rightarrow \nu_{s} $ oscillation as a function of energy and distance from the source and illustrates the fractional oscillation effect normalized to the expected, unoscillated event rate.

The Daya Bay antineutrino detectors are designed to measure the number of $\overline{\nu}_{e}$ interactions as well as the energy of the observed events through the inverse beta decay reaction $\overline{\nu}_e + p \rightarrow e^+ + n$. The total count rate of events as determined from the $e^+ + n$ coincidence as well as the positron energy spectrum are the distinctive event signatures. The neutron thermalizes and captures on Gd followed by a gamma cascade with a total energy of about 8~MeV\@. By using only the Gd-loaded scintillator region in the Daya Bay detectors we can essentially eliminate all possible sources of correlated backgrounds following the inverse beta-decay detection reaction, since the threshold for the
neutron capture on Gd is anticipated to be as high as 6~MeV\@. The Daya Bay detectors are designed to be able to handle singles rates in excess of
1~kHz.

To understand the physics potential of the proposed source experiment we have calculated the predicted number of events and the energy spectrum for two source positions, C and B, in the absence of sterile neutrino oscillation and compared to oscillation into sterile species. For all results shown here we assume a mass splitting of $\Delta m_{41}$=1~eV$^2$ and a mixing angle of sin$^2\theta_{14}$=0.1\@. We analyze the rate and energy of the detected events as a function of baseline from the source. While the antineutrino detectors at Daya Bay can in principle make a measurement of $\theta_{13}$ without position reconstruction of events, position reconstruction is necessary for a sterile neutrino search.  A position resolution of 15~cm is assumed to correlate the reconstructed events with the distance from the $\overline{\nu}_{e}$ source. With moderate position reconstruction the Daya Bay antineutrino detectors will allow us to determine the integral event rate of all energies observed as a function of baseline and the integral event rate of all baselines as a function of energy. We can then deduce the absolute event rate as a function of energy and baseline.

\begin{figure}[htb]
\includegraphics[width=.9\textwidth,angle=0]{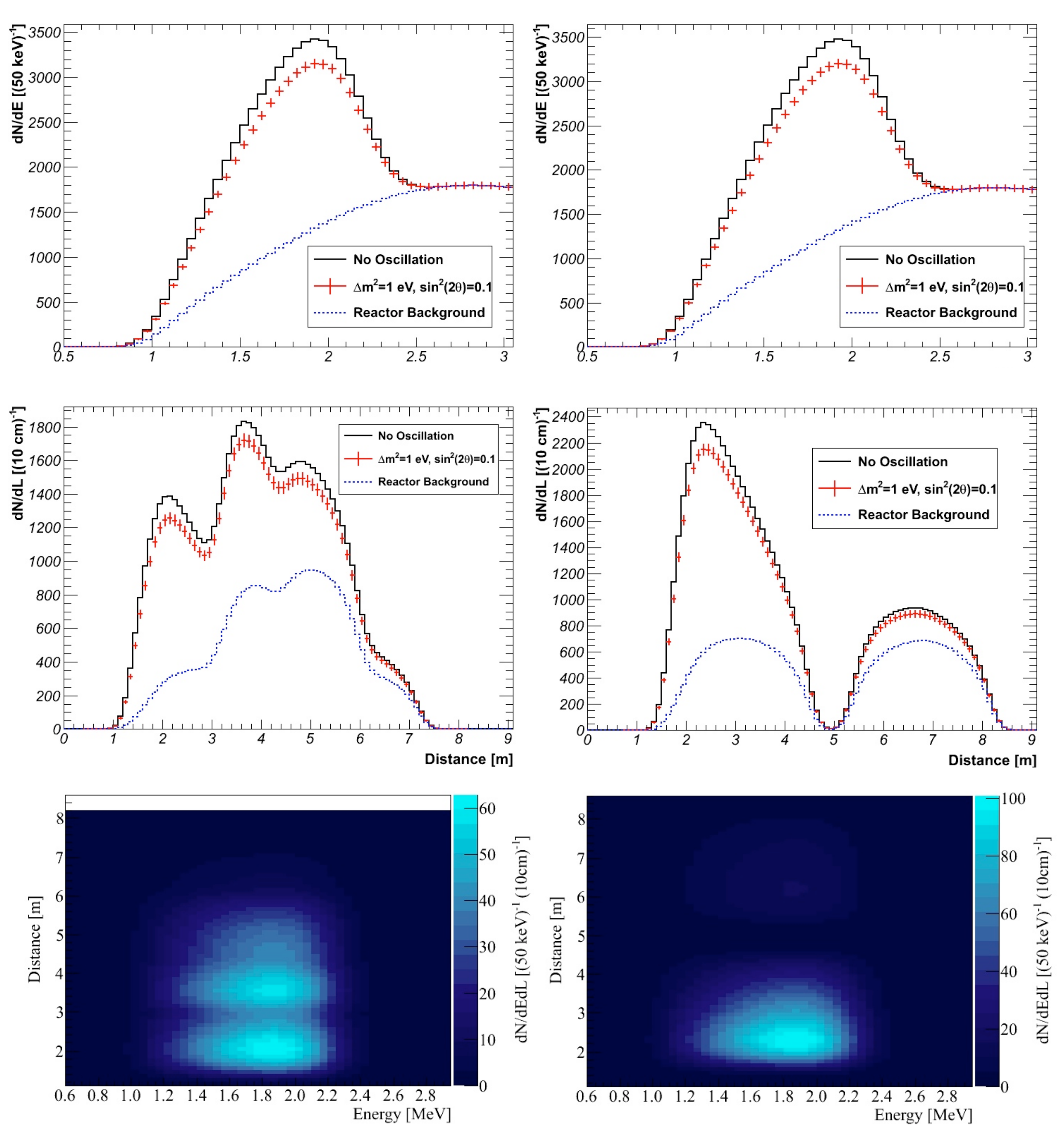}
\caption{Energy and position dependence of the event rates in the antineutrino detectors.  The bottom row shows the 2-dimensional distributions of event rate versus energy and distance from source.  Top and middle rows are the 1-dimensions projections of expected events versus energy (top) and distance from source (middle) for the case of no oscillation (black histogram), the observed event rate in case of $\overline{\nu}_{e} \rightarrow \nu_{s} $ oscillation (red points), and the reactor $\overline{\nu}_{e}$ background (blue dots). Left panels correspond to source position $C$ while right panels are for source position $B$. }
\label{fig:AnalysisPlotfig4}
\end{figure}

Figure~\ref{fig:AnalysisPlotfig4} shows the energy and position dependence of the event rates in the antineutrino detectors for source positions $C$ and $B$.  The bottom row shows the 2-dimensional distributions of event rate versus energy and distance from source.  Top and middle rows are the 1-dimensions projections of expected events versus energy (top) and distance from source (middle) for the case of no oscillation (black histogram), the observed event rate in case of $\overline{\nu}_{e} \rightarrow \nu_{s}$ oscillation (red points), and the reactor $\overline{\nu}_{e}$ background. Left panels correspond to source position $C$ while right panels are for source position $B$. Source position $B$ provides the highest statistics of events as it utilizes the two-fold symmetry of the detector arrangements. With source position $B$ we essentially have a near and far detector arrangements with two primary baselines and two detectors at each baseline. 

\subsection{Backgrounds}

A number of backgrounds will mask the oscillation signal from the proposed source measurement. Correlated and accidental backgrounds to the inverse beta decay signature in the antineutrino detectors are well known to the Daya Bay experiment and will have been studied in great detail before this source measurement is made. In the context of this proposed  sterile neutrino search with a radioactive source we consider only two additional backgrounds: The reactor $\overline{\nu}_{e}$ background from the Daya Bay reactor complex and source-intrinsic backgrounds. 

At distances of about $\sim$1600-2000~m to the detectors, the Daya Bay
and Ling Ao nuclear reactors will be a source of considerable $\overline{\nu}_{e}$ 
background. The size of the reactor background will depend on the energy window under consideration. In the energy range of 1 to 2.2~MeV or 1 to 2.5~MeV respectively of the
positron energy (plus the annihilation $\gamma$), the reactor $\overline{\nu}_{e}$ background will range from $\sim$22,000 events per year (1-2.2~MeV) to 32,000 (1-2.5~MeV) while the
signal from a 500~kCi $^{144}$Ce source will be 31,000 per year for
source position $A$ and $\sim$38,000 and $\sim$37,000 per year for positions $B$ or $C$ respectively.  However, since the shape of the reactor
signal will be well measured and monitored by the detectors near the
reactors, we need to worry mostly only about the statistical fluctuations of the reactor signal. The contributions of the reactor background to the observed signal is shown in Figure~\ref{fig:AnalysisPlotfig4}.

Backgrounds from the antineutrino source itself may also become a concern. While direct $\gamma$ backgrounds can be reduced sufficiently with a W shield around the source, neutrons may become an issue for experiments with sources inside the active detector region. A $^{144}$Ce source can be made by  reprocessing of spent nuclear fuel. Along with the production of $^{144}$Ce,  $^{244}$Cm is produced in the burnup of nuclear reactor fuel~\cite{ORNLdeHart,LLNLHam}. We estimate that one spent nuclear fuel rod contains approximately 100~g of $^{144}$Ce and 5~g of $^{244}$Cm. With half lives of  284.91 days ($^{144}$Ce) and 18.1 years ($^{244}$Cm) respectively and a spontaneous fission  branching fraction of 1.3 $\times$10$^{-6}$ the estimated activities in one spent nuclear fuel rod are
$\sim$11~PBq  ($^{144}$Ce) and 21~MBq ($^{244}$Cm, spontaneous fission only). For a $^{144}$Ce source produced from nuclear fuel, reprocessing would need to have a very high efficiency for rejection of $^{244}$Cm  from $^{144}$Ce.  Even if reprocessing had 100\% efficiency for accepting $^{144}$Ce, and only a 10$^{-6}$ probability for introducing $^{244}$Cm, there would still be a 21~Bq $^{244}$Cm fission rate (for a $^{144}$Ce  source corresponding to one spent nuclear fuel rod).  This would make proposed in-situ source experiments with a $\overline{\nu}_{e}$ source deployed in KamLAND or Borexino challenging, if not impossible. An additional concern is the excess production of high-energy capture gammas due to spent fuel neutrons capturing on detector materials such as stainless steel. This is not such an issue at Daya Bay where we have the unique opportunity to place a source outside the detectors in the water pool with sufficient space for additional shielding around the source. Source-intrinsic backgrounds can be largely mitigated.


\section{Sensitivity of a Sterile Neutrino Search at Daya Bay} 
  
Considering the three source positions $A$, $B$ and $C$ at Daya Bay we have analyzed the sensitivity of the proposed experimental arrangement to the parameters $\Delta m^2_{\rm new}$ and $\sin^2 2\theta_{\rm new}$.  We computed the sensitivity of a source experiment at Daya Bay assuming a 18.5~PBq $^{144}$Ce source corresponding to an intensity of 500~kCi.   The decrease in the antineutrino
source activity over the measurement period of 1 year was estimated to be 66.3\% and taken into account. We found the highest sensitivity for source position $B$ due to the event statistics. For source positions $A$, $B$, and $C$ the total event rates summed over all detectors are $\sim$31,000, 38,000, and 37,000 respectively. In comparison, the reactor background is estimated to be between $\sim$22,000-32,000 events depending on the energy window. See Figure~\ref{fig:AnalysisPlotfig4}. 

The sensitivity was determined using a $\chi^2$ approach.  We assume a 1\% $^{144}$Ce source normalization uncertainty $\sigma_s$, a 1\% reactor normalization uncertainty $\sigma_r$, a 0.5\% detector-to-detector relative uncertainty $\sigma_{AD}$, and a 2\% bin-to-bin uncertainty $\sigma_b$. The statistics corresponds to 1 year of running.  The resulting $\chi^2$ is,
\begin{equation}
\chi^{2} = \sum\limits_{AD}\sum\limits_i\sum\limits_j \frac{\left( N_{obs}^{AD,i,j} - N_{exp}^{AD,i,j}\right)^2}{N_{exp}^{AD,i,j}(1+\sigma_b^2N_{exp}^{AD,i,j})} + \left(\frac{\alpha_s}{\sigma_s}\right)^2 + \left(\frac{\alpha_r}{\sigma_r}\right)^2 + \sum\limits_{AD}\left(\frac{\alpha_{AD}}{\sigma_{AD}}\right)^2.
\end{equation}
The $\chi^2$ first includes a sum over each antineutrino detector.
The indices $i$ and $j$ refer to bins in detected energy and position.
$N_{obs}^{AD,i,j}$ is the number of antineutrino events detected in
each bin, including possible sterile neutrino oscillation.  The
expected number of events assuming no oscillation, $N_{exp}^{AD,i,j}$,
is the sum of events from the $^{144}$Ce source, $S_{exp}^{AD,i,j}$, and
the background from reactor antineutrinos, $R_{exp}^{AD,i,j}$,
\begin{equation}
N_{exp}^{AD,i,j} = \left(1+\alpha_{AD}\right)\left( \left(1+\alpha_{s}\right)S_{exp}^{AD,i,j} + \left(1+\alpha_{r}\right)R_{exp}^{AD,i,j} \right).
\end{equation}
The expected number of events is allowed to vary within the systematic
uncertainties via nuisance parameters; $\alpha_{AD}$ accounts for
efficiency variation between antineutrino detectors, while $\alpha_s$
and $\alpha_r$ account for the $^{144}$Ce source and reactor
normalization uncertainties.

\begin{figure}[htb]
\includegraphics[width=.7\textwidth,angle=0]{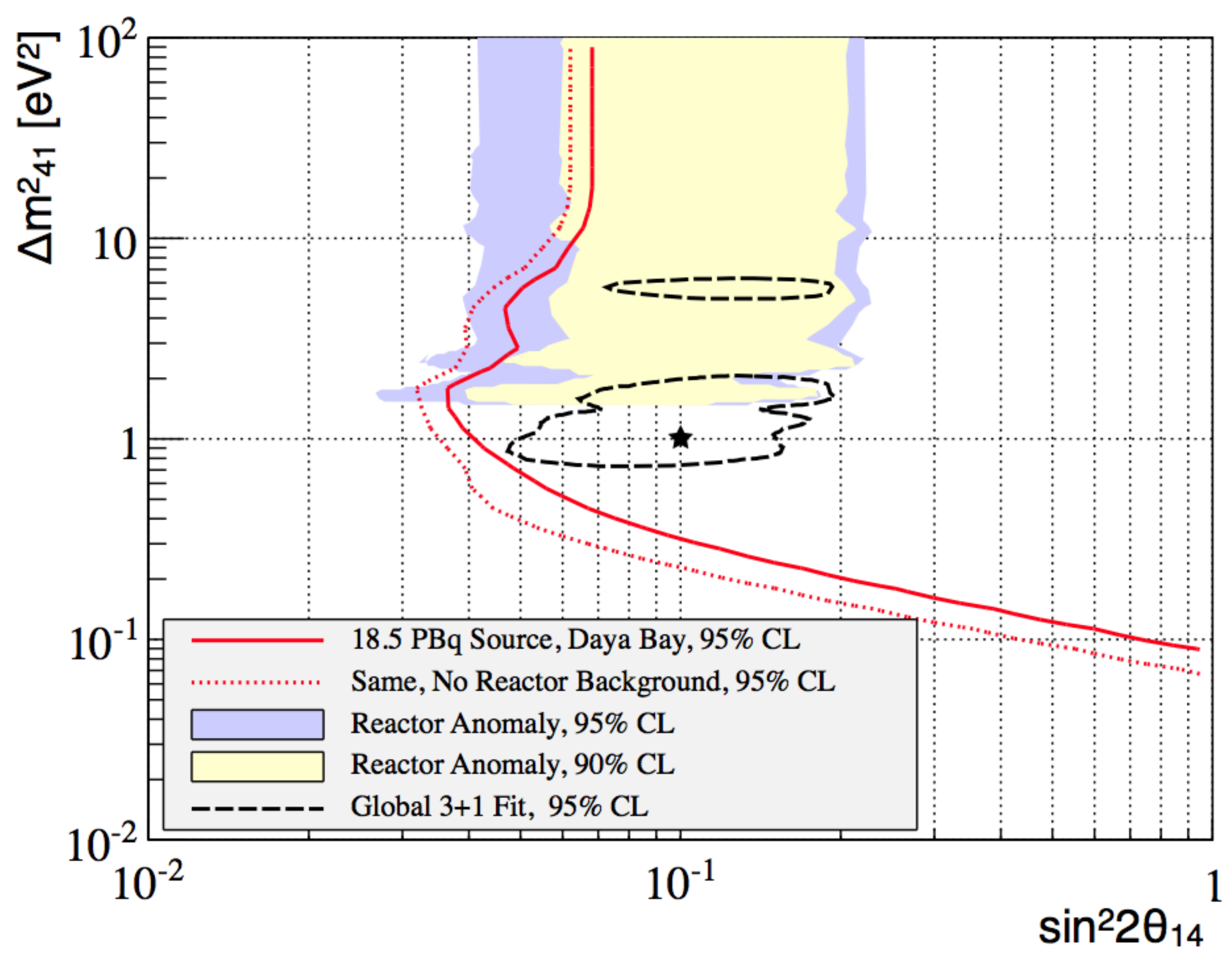}
\caption{Sensitivity of a $\overline{\nu}_{e}$ search at Daya Bay to
  the oscillation parameters $\Delta m^2_{41}$ and sin$^22\theta_{14}$
  assuming a 500~kCi $^{144}$Ce source at position $B$ in the Daya Bay
  Far Hall. We show the 95\% C.L. sensitivity of the Daya Bay source
  experiment with reactor background (red solid) and without (red
  dashed), the 90\% and 95\% C.L. preferred regions of the reactor
  anomaly (shaded yellow and blue)~\cite{Mention}, and the 95\%
  best-fit region from a 3+1 global fit to all neutrino data (dashed
  black)~\cite{Giunti2011}. The parameter space to the left and above
  the Daya Bay sensitivity curve will be excluded at 95\% C.L. The
  star indicates the oscillation parameters $\Delta m^2_{41}$ and
  sin$^22\theta_{14}$ used in Figures~\ref{fig:RateOscfig5} and
  \ref{fig:AnalysisPlotfig4} for the study in this paper.}
\label{fig:Sensfig6}
\end{figure}

For the Daya Bay detector, we assume detector energy and position
resolutions of 9\%/$\sqrt{E(MeV)}$ and 15~cm respectively.  This is
slightly better than the design specifications presented
in~\cite{DayaBay}. A target proton density of 6.4$\times$10$^{28}$
m$^{-3}$ was estimated for the Gd-loaded scintillator. An antineutrino
detector efficiency of 70\% was assumed; dominated by the efficiency
for delayed neutron capture on Gd to produce a signal above
6~MeV\@. The results of our sensitivity calculation for source
position $B$ are shown in Fig.~\ref{fig:Sensfig6} which overlays the
Daya Bay sensitivity to $\Delta m^2_{\rm new}$ and $\sin^2
2\theta_{\rm new}$ with the $\Delta m^2_{\rm 41}$ versus $\sin^2
2\theta_{\rm 14}$ preferred regions of the reactor anomaly and a 3+1
global fit.

The exclusion of the  $\Delta m^2_{41}$ and sin$^22\theta_{14}$  parameter space
 is based on the dependence of the signal on $L/E_{\nu}$,
where $L$ coverage is approximately 1-8~m, and the $E_\nu$ is
between 1.8 and 3~MeV\@.  Both variations of the $\overline{\nu}_{e}$ rate with distance, $L$, and energy,
$E_{\nu}$, are essential. In fact, by considering the ``rate only''
analysis (i.e. integrating over energies and distance $L$) the
exclusion region in $\sin^2 2\theta_{14}$ is reduced
considerably. The energy and position determination of antineutrino
interactions is an important part of the analysis.

The $\overline{\nu}_{e}$ background from the Daya Bay nuclear power plant is an irreducible background to the source $\overline{\nu}_{e}$ signal.  It would be favorable, of course, to build a source experiment with antineutrino detectors in an underground location far from nuclear reactors; but this may not be feasible. The advantage of the proposed source experiment at Daya Bay is the existence of multiple antineutrino detectors in an underground water pool with the access and infrastructure needed for a source deployment.  Our calculations show that in the absence of a reactor $\overline{\nu}_{e}$ background and with a 500~kCi source we would
reach $\sin^2 2\theta_{14}\ge$0.06 compared with
$\sin^2 2\theta_{14}\ge$0.07 for the case with the reactor background (see Fig.~\ref{fig:Sensfig6}). At half of the source strength the sensitivity in the presence of the $\overline{\nu}_{e}$ reactor background would decrease from $\sin^2 2\theta_{14}\ge$0.07 to about $\sin^2 2\theta_{14}\ge$0.08. 

A more detailed analysis, beyond the simple approach presented here,
may reveal an improved sensitivity to sterile neutrinos. Over the coming years the Daya Bay experiment will make a high-precision measurement of  $\theta_{13}$. During this measurement Daya Bay is expected to
constrain the reactor neutrino interaction rate to sub-percent
precision and provide a measurement of the detector-to-detector relative uncertainty. The relative rate of reactor backgrounds will provide a
strong constraint on the relative efficiency of the four antineutrino
detectors. The search for sterile neutrinos with a $\overline{\nu}_{e}$ source at Daya Bay is incompatible with the measurement of $\theta_{13}$ and will have to take place after the measurement of $\theta_{13}$ is completed. 

The multiple antineutrino detectors at the Daya Bay Far Hall provide a unique laboratory for the measurement of antineutrinos. Exploiting the symmetric positions of the detectors relative to the $^{144}$Ce source will provide further constraints on
systematic uncertainties, and placing a source at multiple locations could improve the sensitivity further.

\section{Conclusions}

Sterile neutrino oscillations with mass of $\geq$1~eV can be tested
using a 500~kCi $^{144}$Ce-$^{144}$Pr antineutrino source in the Far
Hall of the Daya Bay reactor experiment.  With one year of data, the
allowed parameter region given by the recent ``reactor anomaly'' and
global 3+1 fits can be tested.  The Daya Bay Far Hall offers
significant advantages for the placement of the source, its cooling,
and the shielding of source-intrinsic backgrounds. One of the major
technical advantages is the placement of the source outside the Daya
Bay antineutrino detectors. The geometric arrangement of the four
identical Daya Bay detectors and the flexibility to place the
$\overline{\nu}_{e}$ source at multiple locations inside the water
pool allows for additional control of experimental systematics.

\section*{Acknowledgments}

Funding for this work was provided in part by DOE Office of High-Energy Physics under Contract No. DE-FG02-95ER40896, by the US DOE Contract No. DE-FG02-88ER40397, NSF grant 0555674, the Alfred P. Sloan Foundation, and the University of Wisconsin Foundation. We thank T.~Lassere and B.~Balantekin for useful comments and discussions. 


\end{document}